\documentclass{myws-p8}
\usepackage{graphicx,psfrag,amsmath}

\newcommand{\m}{m_{\mbox{\tiny$ N$}}}
\newcommand{\M}{M_{ \pi}}

\newcommand{\F}{F_\pi}

\begin{document}

\title{Lorentz Invariant Baryon CHPT}

\author{T. Becher\footnote{\uppercase{S}\lowercase{upported by the }\uppercase{S}\lowercase{wiss }\uppercase{N}\lowercase{ational }\uppercase{S}\lowercase{cience }\uppercase{F}\lowercase{oundation.}\uppercase{ T}\lowercase{he work presented in this article was done in collaboration with }\uppercase{H.~L}\lowercase{eutwyler. }}}

\address{Newman Laboratory of Nuclear Studies\\       
Cornell University,\\ Ithaca, NY 14853-5001,
USA\\E-mail: becher@mail.lns.cornell.edu}

\maketitle

\abstracts{ Using the example of the elastic $\pi N$-amplitude, we
discuss the low energy expansion of QCD amplitudes in the sector with
baryon number one. We show that the chiral expansion of these
amplitudes breaks down in certain regions of phase space and present a
framework which leads to a coherent description throughout the low
energy region, while keeping Lorentz and chiral invariance manifest at
every stage of the calculation. We explain how to construct a
representation of the pion nucleon scattering amplitude in terms of
functions of a single variable, which is valid to $O(q^4)$ and properly
accounts for the $\pi\pi$- and $\pi N$-cuts required by unitarity. }

\section{Introduction}
The chiral symmetry, which the Lagrangian of QCD reveals in the limit
of vanishing quark masses, can systematically be incorporated into an
effective Lagrangian describing the interaction of the lowest lying
baryons and mesons. The effective description is based on a
simultaneous expansion of the QCD Greens functions in powers of the
light quark masses and the momenta of the Goldstone bosons. 

In the vacuum sector of the theory, where the only low energy
singularities arise from the propagation of the Goldstone bosons,
dimensional regularization yields homogeneous functions of the
Goldstone momenta and masses, so that each graph has an unambiguous
order in the chiral expansion. The singularity structure in the sector
with baryon number one is more complicated. The expansion of
the corresponding amplitudes in the momenta and masses of the Goldstone
bosons leads to an expansion of the nucleon propagator around the
static limit. This expansion of the nucleon propagators is implemented
ab initio in the framework called heavy baryon chiral perturbation
theory\cite{hbchpt} (HBCHPT). However, as we will show in section
\ref{sec:lowEsing}, it fails to converge in part of the low energy
region. The breakdown is related to the fact that the expansion of
the nucleon propagator ruins in some cases the singularity structure
of the amplitudes. This makes it desirable to perform the calculations
in a relativistic framework. In doing so, the correct analytic
properties of the amplitudes are guaranteed and one can address the
question of their chiral expansion in a controlled way.

In the relativistic formulation of the effective theory a technical
complication arises from the fact that in a standard regularization
prescription, like dimensional regularization, the low energy
expansion of the loop graphs starts in general at the same order as
the corresponding tree diagrams\cite{gss}. Since the contributions
that upset the organization of the perturbation expansion stem from
the integration region of large loop momentum of the order of the
nucleon mass, they are free of infrared singularities. In section
\ref{sec:infrared} we show that the infrared singular part of the loop
integrals can be unambiguously separated from the remainder, whose low
energy expansion to any finite order is a polynomial in the momenta
and quark masses. Moreover the infrared singular and regular parts of
the amplitudes separately obey the Ward identities of chiral
symmetry. This ensures that a suitable renormalization of the
effective coupling constants removes the infrared regular part
altogether, so that we may drop the regular part of the loop integrals
and {\em redefine} them as the infrared singular part of the integrals
in dimensional regularization, a procedure referred to as {\em
infrared regularization}\cite{Becher:1999he}. The representation of
the various quantities of interest obtained in this way combines the
virtues of HBCHPT and the relativistic formulation: Both the chiral
counting rules and Lorentz invariance and are manifest at every stage
of the calculation.

In the meantime this framework has been used to calculate the
scalar\cite{Becher:1999he}, axial\cite{julia} and electro-magnetic
formfactors\cite{Kubis} as well as the elastic pion nucleon
amplitude\cite{BL} to fourth order in the chiral
expansion.\footnote{See also B.~Kubis contribution to these
proceedings} In section \ref{sec:piN}, we discuss the structure and
the low energy expansion of the pion nucleon amplitude. To fourth
order in the chiral expansion the amplitude can be given in terms
functions of a single variable, either $s$, $t$ or $u$.

\section{Low energy singularities of one nucleon amplitudes\label{sec:lowEsing}}
\begin{figure}[h]
\begin{center}
\begin{tabular}{ccccccccc}
 \raisebox{0.0\textwidth}{\includegraphics[width=0.12\textwidth]{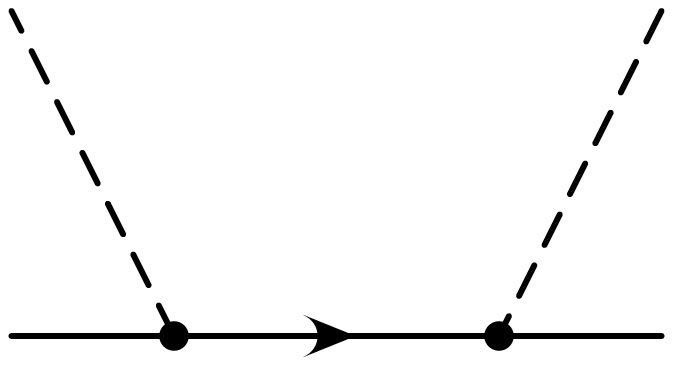}} & &
  \raisebox{0.0\textwidth}{\includegraphics[width=0.12\textwidth]{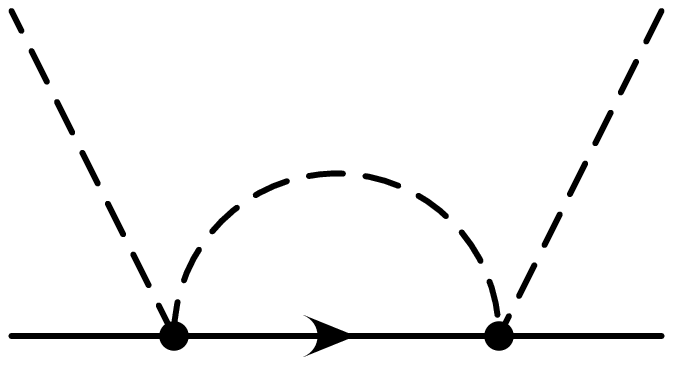}} &&
  \raisebox{0.0\textwidth}{\includegraphics[width=0.12\textwidth]{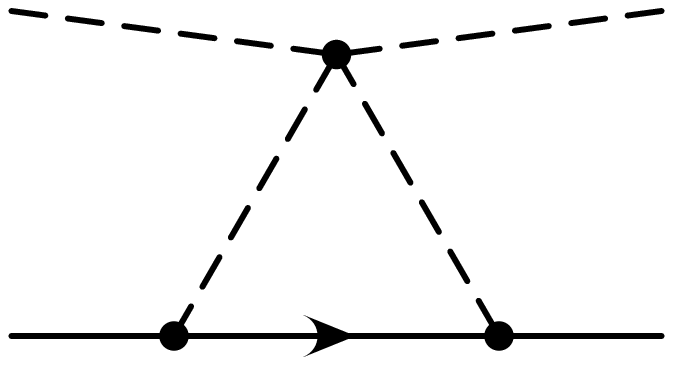}} &&  
\raisebox{0.0\textwidth}{\includegraphics[width=0.12\textwidth]{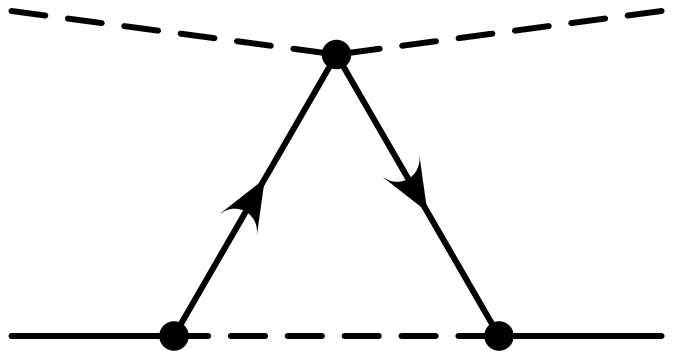}} &&
  \raisebox{0.0\textwidth}{\includegraphics[width=0.12\textwidth]{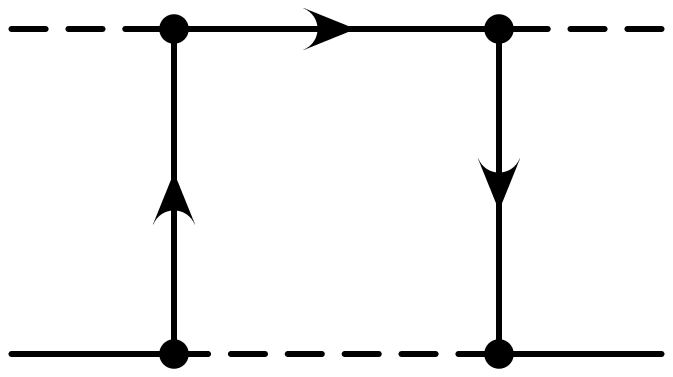}}
\\ && && && \\
\small (a) &&\small (b) &&\small (c) &&\small (d)&&\small (e)
\end{tabular}
\end{center}
    \caption{\label{fig:sing} Examples of contributions to the $\pi N$ scattering amplitude with different singularity structure.}
\end{figure}
The low energy singularities of the one nucleon amplitudes are due to
the propagation of pions and nucleons. In the example of the pion
nucleon amplitude these include poles at $s=\m^2$ and $u=\m^2$ due to
one nucleon exchange, branch points at $s=(\M+\m)^2$ and $u=(\M+\m)^2$
due to $\pi N$ intermediate states as well as a branch point at
$t=4\M^2$ due to $\pi \pi$-intermediate states in the $t$-channel (see
figure \ref{fig:sing}). Note that some of the graphs, like figs.~\ref{fig:sing}d and \ref{fig:sing}e, also involve a cut
for $t>4\m^2$ due to the presence of $\bar{N}N$ intermediate states in
the $t$-channel. This singularity lies however outside the low energy
region and is absent to any finite order of the chiral
expansion.

In performing the low energy expansion we must choose the kinematic
variables in such a way that the singularities of the amplitude stay
put in order to avoid a breakdown of the expansion near the
singularity. We will illustrate this first with the trivial example of
the nucleon Born term and then discuss the singularity structure of
the triangle graph, figure \ref{fig:sing}c. In addition to the cut for
$t>4\M^2$, this graph has a branch point at
\begin{equation*}
t=4\M^2-\frac{\M^4}{\m^2}
\end{equation*}
on the second Riemann sheet. Since the difference between the two
singularities is of higher order, they will coalesce if one performs
the chiral expansion, causing a breakdown of the expansion in the
vicinity of the threshold at $t=4\M^2$.

\subsection{Born term}

Denoting the momentum of the incoming nucleon and pion with $P$
resp.~$q$, the momentum dependence of the $s$-channel Born
contribution to the amplitude, fig.~\ref{fig:sing}a is given by
\begin{equation}\label{eq:born}
T_{fi} \propto \frac{\bar{u}'\,q\!\!\!/ (P\!\!\!\!/+q\!\!\!/+\m)\,q\!\!\!/\, u}{\m^2-(P+q)^2} \, .
\end{equation}
If we expand in the momentum of the incoming pion, the propagator denominator gets replaced by the geometric series
\begin{equation}\label{eq:bornex}
\frac{1}{\m^2-(P+q)^2}\rightarrow-\frac{1}{2\,P\cdot q}\left\{1-\frac{\M^2}{2P\cdot q}+\ldots\right\} \, .
\end{equation}
The location of the singularity has moved to $P\cdot q=0$ or
$s=\m^2+\M^2$ and the expansion breaks down for $|P\cdot
q|<\M^2/2$. Clearly, this problem can be cured by performing the
expansion in $q$ at fixed $\tilde{P}=P+q/2$ instead of fixed $P$. Note
however that we need to choose appropriate variables for each graph
separately in order not to spoil its specific singularity
structure. In the example of the graph shown in fig.~\ref{fig:sing}b the expansion at fixed $P$ is convergent, while the expansion at fixed $\tilde{P}$ breaks down in the vicinity of the threshold at $P\cdot q=\m \M$.

\subsection{Triangle graph}

The discontinuity of the triangle graph, fig.~\ref{fig:sing}, can be obtained from the extended unitarity relation
\begin{equation}\label{t-channel unitarity}
\mbox{Im}\,f^J_\pm(t)=\left\{1-4M_\pi^2/t\right\}^{\frac{1}{2}}\;
t_{J}^{I}(t)^\star\; 
f^J_\pm(t)\, ,\;\;\;\;4\M^2<t<16\M^2 \,.
\end{equation}
The quantities $f^J_\pm(t)$ represent the $t$-channel partial waves of
the $\pi N$ scattering amplitude. The quantum number $J$ stands for
the total angular momentum, while the lower index refers to the spin
configuration . There is no need to in addition specify the isospin
quantum number, because it is determined by the total angular
momentum: $J$ even $\rightarrow I=0$, $J$ odd $\rightarrow I=1$. The
functions $t_{J}^{I}(t)$ denote the
partial wave projections of the $\pi\pi$ scattering amplitude. To
order $O(q^4)$ the evaluation of this condition is particularly
simple, because the $\pi\pi$ scattering amplitude is needed only to
leading order, where it is a polynomial.  Hence only the partial waves
with $J=0,1$ contribute.

To obtain the imaginary part of the triangle graph, we insert the lowest partial waves of the nucleon Born term into the unitarity relation (\ref{t-channel unitarity}). For $J=0$ the pertinent partial waves are
\begin{align}
{f_B}_+^0(t)&=\frac{g_{\pi N}^2\,\m}{4\pi}\left\{ f(\kappa)-\frac{t}{4\,m^2}\right\} &\text{and} && t_0^0&=\frac{2\,t-\M^2}{32\pi\F^2} \nonumber\, .
\end{align}
The function $f(\kappa)$ stands for
\begin{align*}
f(\kappa)&=\frac{\arctan\kappa}{\kappa}\, ,& \kappa
 &=\frac{\sqrt{(t-4\,\M^2)\,(4\,\m^2-t)}}{t-2\,\M^2}\, .\nonumber
\end{align*}
The problem addressed above shows up in this expression: The quantity
$\kappa$ represents a term of order $O(1/q)$ and the standard chiral
expansion of $f(\kappa)$ corresponds to an expansion of
$\arctan\kappa$ around $\kappa=\infty$. This expansion only converges
for $|\kappa|>1$. In the vicinity of the threshold at $t>4\M^2$ this
condition is not met and the chiral expansion diverges. The breakdown
of the expansion is related to the fact that $\arctan\kappa$ has a
branch points at $\kappa=\pm i$. To $O(q^4)$, the imaginary part of
the invariant amplitude $D^+$ (see eq.~(\ref{eq:amp}) for its
definition) can be obtained from
\begin{equation*}
\mbox{Im}_t D^+ =\frac{4\,\pi}{\m^2}\, \mbox{Im} f_+^0 +O(q^5) \, .
\end{equation*}
The analytic continuation of the
scattering amplitude $D^+$ to the second Riemann sheet therefore contains a
branch point at
\begin{equation*}
\frac{(t-4\,\M^2)\,(4\,\m^2-t)}{(t-2\,\M^2)^2}=-1 \rightarrow t=4\M^2-\frac{\M^4}{\m^2} \, .
\end{equation*}

\section{Infrared regularization\label{sec:infrared}}
Using the example of the self energy integral, we explain how to
separate the infrared singular part of the loop integrals from their
regular part. As is evident from eqns.~(\ref{eq:born}) and
(\ref{eq:bornex}), the nucleon propagator counts as a quantity of
$O(1/q)$ at tree level. While the infrared regular part of the loop integrals
violates this counting, we show in the following that the infrared singular part respects the counting rules established at tree level.
\subsection{Self-energy integral}
\begin{figure}[h]\centering\psfrag{P}{\footnotesize $P$}\psfrag{k}{\footnotesize $k$}\psfrag{P-k}[B]{\footnotesize $\phantom{k}P-k$}
\begin{tabular}{ccc}
 \raisebox{-0.008\textheight}{\includegraphics[width=0.26\textwidth]{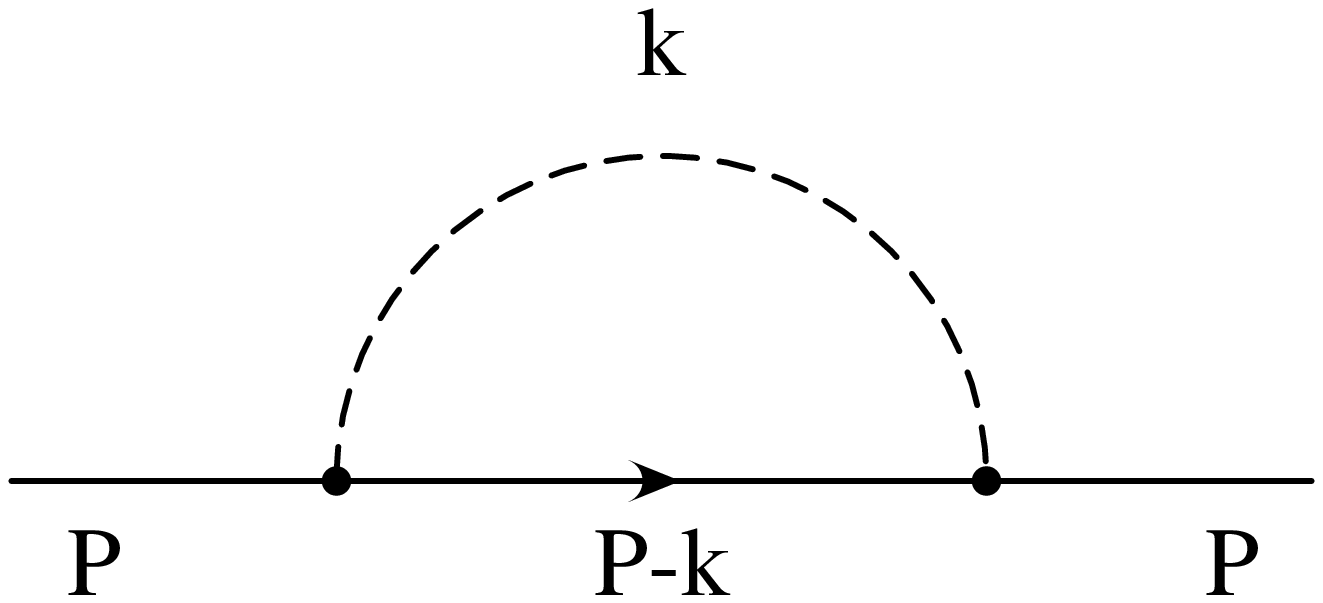}}
& \hspace*{0.14\textwidth} &
\psfrag{k}{}\psfrag{P}{}\psfrag{P'}{}
\includegraphics[width=0.2\textwidth]{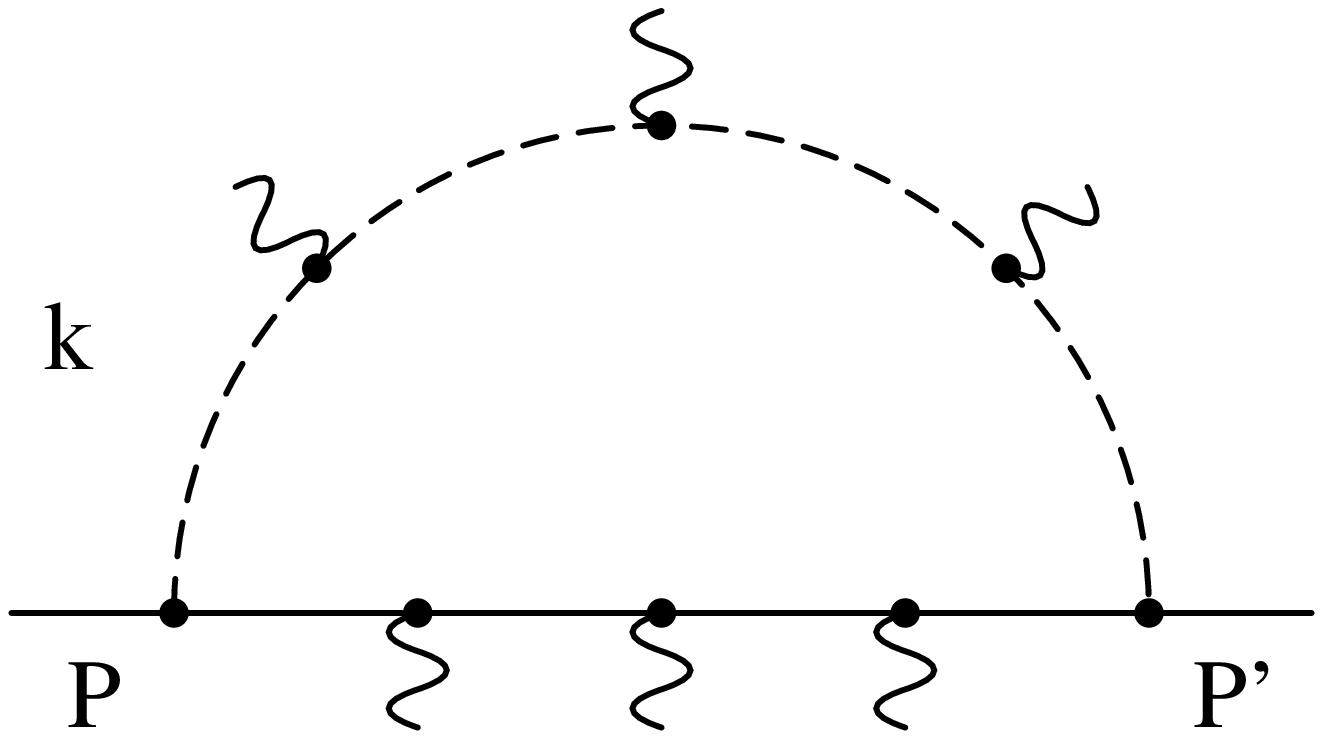}\\
\small (a)  &   & \small(b)
\end{tabular} 
\caption{\label{fig:oneloop} a.) Self energy graph. b.) Form of the general one loop graph.}
\end{figure}

The simplest example of a one loop contribution, the self energy graph, is shown in figure \ref{fig:oneloop}a.
 The corresponding scalar loop integral has the form
\begin{align}\label{eq:oneloop}
H(P^2)&=\frac{1}{i}\int\frac{d^d k}{(2\pi)^d}
\;\frac{1}{\M^2-k^2}\;\frac{1}{\m^2-(P-k)^2}
=\frac{1}{i}\int\frac{d^d k}{(2\pi)^d}\;\frac{1}{a}\;\frac{1}{b}  \nonumber\, .
\end{align}
We need to analyze the integral for external momenta in the vicinity
of the mass shell: $P=\m v+r$, where $v$ is a time-like unit vector
and $r$ is a quantity of order $q$.  In the limit $M\rightarrow 0$,
the integral develops an infrared singularity, generated by small
values of the variable of integration, $k=O(q)$. In that region, the
first factor in the denominator is of $O(q^2)$, while the second is of
order $O(q)$. Accordingly the chiral expansion of the integral
contains terms of order $q^{d-3}$. The order of the infrared singular
part follows from the counting rules at tree level, because it is
generated by the region of integration, where the momenta flowing
through the propagators are of the same order as in the tree level
graphs.  The remainder of the integration region does not contain
infrared singularities and may thus be expanded in an ordinary Taylor
series. An evaluation of the integral at its threshold at
$P^2=s_+=(\m+\M)^2$ nicely shows the two parts
\begin{equation}\label{intthreshold}
H(s_+)=\frac{\Gamma(2-\frac{d}{2})}
{(4\pi)^\frac{d}{2}(d-3)}\;\Big\{\frac{\M^{d-3}}{(\m+\M)}+
\frac{\m^{d-3}}{(\m+\M)}\Big\}=I+R  \, .
\end{equation}
The infrared singular part $I$ is proportional to $\M^{d-3}$, while the
remainder $R$ is proportional to $\m^{d-3}$ and does therefore not contain a
singularity at $\M= 0$, irrespective of the value of $d$. Since the regular
contribution stems from a region where the variable of integration $k$ is of
order $O(\m)$, it violates the tree-level counting rules.

\subsection{Relation to HBCHPT}
The infrared singular part of the loop integrals can be obtained
exchanging the order of the chiral expansion and the loop
integration. To perform the expansion, both the residual momentum
$r_\mu=P_\mu-\m v_\mu$ and the loop momentum $k$ are counted as
$O(q)$. Each term in the expansion of the integrand has a trivial
dependence on the nucleon mass $\m$ and does therefore not contain a
regular part. Performing the loop integration one obtains the infrared part of the integral by summing the individual terms of the series.

\begin{figure}
\begin{center}
\begin{align}
\includegraphics[width=0.15\textwidth]{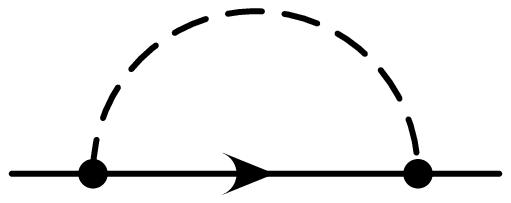} && \rightarrow&& \includegraphics[width=0.15\textwidth]{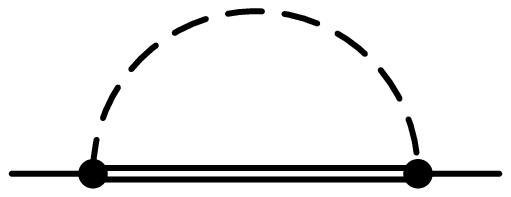} &&+&&  \raisebox{-0.006\textwidth}{\includegraphics[width=0.15\textwidth]{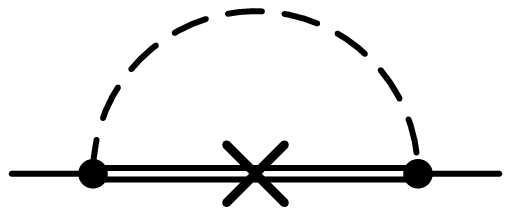}}&& +&& \ldots \nonumber
\end{align}
\label{fig:ins}
\caption{Internal line insertions. The double line denotes the heavy
      baryon propagator $[2\m\, v\cdot( k-r)-i\epsilon]^{-1}$, the
      cross an insertion of $(k-r)^2$.}
\end{center}
\end{figure}

Let us illustrate this again for the self energy integral. In this
case the expansion amounts to replacing
\begin{align}
 \frac{1}{\m^2-(P-k)^2} && \rightarrow && \frac{1}{2
  \m\, v\cdot( k-r)} &&+ &&\frac{(k-r)^2}{\{2 \m\, v\cdot(
  k-r)\}^2} && + && \ldots\nonumber 
\end{align}
and the relativistic self-energy integral gets replaced by a sum of integrals which arise in HBCHPT (see fig.~\ref{fig:ins}):
\begin{align}
H\rightarrow I& =\sum_{n=1}^{\infty}I_n \, , & I_n=\int_k \frac{1}{\M^2-k^2}\frac{(k-r)^{2(n-1)}}{[2\m(v\cdot k-v\cdot r)]^n}\nonumber\, .
\end{align}
Let us evaluate the integrals at $P^2=s_+=(\M+\m)^2$ by setting $r_\mu=\M v_\mu$.
After expressing $(k-r)^2=(k-\M v)^2$ in terms of the denominators of the integrals and by using the fact that closed loops of static propagators vanish, we obtain the relation 
\begin{equation}
I_n(s_+)=-\frac{\M}{\m}\,I_{n-1}(s_+) \, .
\end{equation}
Since we have expanded the nucleon propagator, the integrals $I_n$
depend only trivially on $\m$ (they don't have a regular part) and
dimensional analysis gives $I_1=c(d) \M^{d-3}/\m$. Therefore
\begin{equation*}
I(s_+)=\sum_{n=1}^{\infty} I_n(s_+)=\sum_{n=1}^{\infty} (-)^n\frac{\M^n}{\m^n} I_1(s_+)=c(d)\frac{\M^{d-3}}{\m+\M} \, .
\end{equation*}
An explicit calculation of $I_1(s_+)$ shows that the constant $c(d)$ is indeed the same as in (\ref{intthreshold}).

In the case of the self-energy integral it is possible to sum the
kinematic insertions explicitly \cite{Becher:1999he,Tang}, but for a
general integral, the above procedure would be rather clumsy. It
however illustrates the relationship between HBCHPT and infrared
regularization: The HBCHPT representation of the one nucleon
amplitudes is obtained by first performing the chiral expansion and
then the loop integration. The infrared regularized amplitudes can be
obtained from HBCHPT by resumming the graphs with kinematic
insertions.

\subsection{Feynman parameterization}
The explicit expression for the loop integral $H(P^2)$ in $d$ dimensions is
lengthy, but the splitting into the infrared singular part $I$ and regular
part $R$ is easily obtained in the Schwinger-Feynman representation of the
integral. 
\begin{align}
  H\;&=\;\int\frac{d^d k}{(2\pi)^d}\, \frac{1}{a}\,\frac{1}{b} = \int_0^1
  {dz}\,\int\frac{d^d k}{(2\pi)^d}\,\frac{1}{[\,(1-z)\, a+z
    \,b\,]^2}\nonumber\\ 
  &=\;\int_0^{\infty}-\int_1^{\infty}dz\,\int\frac{d^d k}{(2\pi)^d}\,
  \frac{1}{[\,(1-z)\, a+z\, b\,]^2}\; 
  = I+R \,.\label{eq:feynman}
\end{align}
The case of a general one loop integral (see fig.\ref{fig:oneloop}b)
can be reduced to the self-energy integral, by first using the
ordinary Feynman parameterization to combine all the meson propagators
into a single one. After combining also the nucleon
propagators in the same way, the general integral can be written as
an integral over the self-energy integral $H$ and derivatives with
respect to its masses. Its infrared part is then obtained by replacing
$H$ by the Feynman parameterization for $I$ given in
(\ref{eq:feynman}). Feynman parameterizations of the infrared singular
and regular parts of a general one loop integral can be found in the literature
\cite{Becher:1999he}.

\subsection{Chiral symmetry}
In the framework of the relativistic effective theory, the evaluation
of an amplitude to one loop yields three contributions, arising from
(a) the tree graphs, (b) the infrared singular and (c) the infrared
regular part of the one loop graphs. At a given order in the chiral
expansion, the one particle irreducible components of the regular part
are polynomial in the external momenta. In coordinate space this
contribution is given by local terms and thus equivalent to the tree
graphs generated by a suitable Lagrangian $\Delta {\cal L}$. So that, if we replace the effective Lagrangian by
\begin{equation*}
{\cal L}_{\text{eff}}'={\cal L}_{\text{eff}}+\Delta {\cal L} \, ,
\end{equation*}
we may drop the regular part of the loop integrals.

We show now that the modified Lagrangian ${\cal L}_{\text{eff}}'$
occurring in our framework respects the Ward identities of chiral
symmetry. In the limit of vanishing quark masses these identities
interrelate the Green's functions of the form
\begin{equation*}
<N(P',s')|{\bf T}\left\{ O_1(x_1) \dots O_n(x_n)\right\}|N(P,s)> \, ,
\end{equation*}
where the operators $O_i$ stand for vector, axial-vector, scalar and pseudo-scalar quark-currents.

Since ${\cal L}_{\text{eff}}$ is chirally symmetric, the tree level
amplitudes (a) respect the Ward identities. Because dimensional
regularization preserves the symmetry of ${\cal L}_{\text{eff}}$, the
same is true for the loop contribution (b)+(c) for {\em arbitrary}
values of the regularization parameter $d$. Since (b) involves only
fractional and (c) only integer powers of the chiral expansion
parameter, the sum of the two can only fulfill the Ward identities if they
are separately fulfilled by the two parts and we conclude that the
Lagrangian $\Delta {\cal L}$, which generates the contributions (c) is
indeed chirally symmetric.

\section{Representation of the $\pi N$-scattering amplitude at $O(q^4)$\label{sec:piN}}
\begin{figure}[h]\centering
\psfrag{q,a}[rb]{\footnotesize$q,a$}
\psfrag{q',a'}[lb]{\footnotesize$q',a'$}
\psfrag{p}[rt]{\footnotesize$P$}
\psfrag{p'}[lt]{\footnotesize$P'$}
\psfrag{s}[b]{\footnotesize$s$}
\psfrag{t}[]{\footnotesize$t$}
 \includegraphics[width=1in]{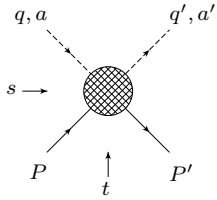}
    \caption{\label{cap:kin}Kinematics of the elastic $\pi N$ scattering
    amplitude $P$, $q$ ($P'$,$q'$) denote the momentum of the incoming
    (outgoing) nucleons and pions, and $a$ ($a'$) stands for the isospin
    indices of the incoming (outgoing) pion. }
  \end{figure}

The virtues of CHPT are twofold: One one hand chiral symmetry
interrelates different amplitudes, on the other hand the framework
allows one to obtain a parameterization of a given amplitude in a
number of low energy parameters not fixed by the symmetry.

In the following we focus on the energy dependence of the elastic pion
nucleon scattering amplitude in the isospin limit, obtained on the
basis of a calculation in infrared regularization to $O(q^4)$.\cite{BL} We
explain, how one arrives at a representation of the amplitude in
terms of functions of a single variable without invoking an expansion
of its infrared singularities. As discussed in section
\ref{sec:lowEsing}, the chiral expansion of the triangle graph does
not converge around its threshold. The expansion of the other graphs
is convergent, but the convergence is in general rather slow and
depends on the choice of the kinematic variables, which are kept fixed
while one performs the expansion.

In section \ref{sec:lowEsing}, we have listed some of the analytic
properties of the scattering amplitude. In the chiral representation
of the amplitude, the corresponding cuts are described by loop
integrals. These loop integrals in general also contain singularities
outside the low energy region and we will in the following simplify the
representation of the amplitude by chirally expanding the contributions from
these singularities. As a result of these simplifications,
we can represent the amplitude as the sum of the Born term, a
polynomial containing the low energy constants and nine functions of a
single variable, which are given in terms of integrals over the
imaginary parts of the loop integrals.

In the context of our low energy analysis, the standard decomposition
of the scattering amplitude into invariant amplitudes $A$ and $B$ is
not well suited, because their leading order contributions
cancel. Replacing $A$ by $D=A+\nu B$, with $\nu=(s-u)/4\m$, the
scattering amplitude is given by
\begin{equation} T^{\pm}=\bar{u}'\left\{ D^{\pm}
  -\frac{1}{4 \m} [\,q\hspace{-0.5em}/^{\,\prime},\,
  q\hspace{-0.5em}/\hspace{0.15em}]B^{\pm} \right\}u \,
  , \label{eq:amp}
\end{equation}
where the upper index refers to the isospin decomposition
\begin{equation*}
T_{a'a} = \delta_{a'a}T^+ + 
\mbox{$\frac{1}{2}$}[{\bf \tau}_{a'},{\bf \tau}_a]\,T^- \, .
\end{equation*}
The evaluation of the chiral perturbation series to one loop allows us
  to calculate the amplitudes $D^\pm$ and $B^\pm$ to $O(q^4)$ and
  $O(q^2)$, respectively.

\subsection{Simplification of the $t$-dependence of the amplitude\label{sec:simp}}

The triangle graph with two nucleon propagators, fig.~\ref{fig:sing}d,
and the box graph, fig.~\ref{fig:sing}e, have a cut for
$t>4\m^2$. Because this cut lies outside the low energy region, we can
expand the corresponding graphs in the variable $t$. The higher order
contributions in $t$ are suppressed as $t/4\m^2$ and are therefore
also of higher order in the chiral expansion. After this
simplification, the amplitude can be expressed in terms of functions
of a single variable: The contribution of the graphs with an
$s$-channel cut is a linear function of $t$ and the contribution of the graphs
with a $t$-channel cut is linear in $\nu$.

\subsection{Dispersive representation}

By construction, the infrared regular part of the loop integrals has
no singularities in the low energy region. It however has unphysical
singularities outside the low energy region\cite{Becher:1999he} and
since the infrared part is given by the difference between the full
integrals and their regular part, these singularities are also present
in the infrared part. The infrared part of the self energy integral,
for example, has a cut for $s<0$ and a pole at $s=0$. Also these
singularities can safely be chirally expanded. Let us illustrate in
the example of the forward amplitude $\bar{D}^+(\nu,t=0)$ how this can
be achieved (we use the bar to indicate that the pseudo-vector Born
term has been subtracted from $D^+$). In a first step, we expand the
amplitude around $\nu=0$ to $O(\nu^4)$:
\begin{equation*}
\bar{D}^+(\nu,0)=d_0 + d_1 \nu^2 + d_2 \nu^4
+ O(\nu^6)
\end{equation*}
The absence of odd powers of $\nu$ is due to crossing symmetry. Note
that we have {\em not} expanded the coefficients $d_i$ in powers of
the quark masses, because this expansion converges only slowly, especially in the case of $d_2$. Once its contributions to the polynomial part are
subtracted, the remaining contribution from the physical $s$- and
$u$-channel cut is given by the dispersion integral
\begin{equation*}
\hat{D}^+(\nu)=\frac{2}{\pi}\int_{\M}^\infty \frac{\mbox{d}\nu'}{{\nu'}^2-\nu^2}\frac{\nu^6}{{\nu'}^5}\,\mbox{Im}_s D^+(\nu,0) \, .
\end{equation*}
$\mbox{Im}_s D^+(\nu,0)$ denotes the discontinuity
over the right hand cut of the $s$-channel graphs, which is a consequence of
unitarity and is the same in dimensional and infrared regularization.

If we subtract from the amplitude both the polynomial and the
dispersion integral, we are left with
\begin{equation}
D^+_{\text{reg}}(\nu)=\bar{D}^+(\nu,0)-\hat{D}^+(\nu)- d_0 - d_1 \nu^2 - d_2 \nu^4 \, ,
\end{equation}
which receives contributions only from the singularities outside the low energy region. Since we have accounted for all infrared singularities and
since $D^+_{\text{reg}}(\nu,0)$ is proportional to $\nu^6$, its chiral
expansion involves only terms of order $O(q^6)$ and higher and we can
drop $D^+_{reg}$ to the accuracy of our calculation.The representation
of the forward amplitude now takes the form
\begin{equation}
\bar{D}^+(\nu,0)=d_0+d_1 \nu^2+d_2 \nu^4+ \hat{D}^+(\nu) + O(p^4) \, .
\end{equation}
The coefficients $d_i$ contain low energy constants, whose values are
not fixed by the symmetry. By construction $\hat{D}^+(\nu)$ does not contribute to the coefficients $d_i$, so that we can identify these constants with the lowest coefficients of the subthreshold-expansion of H\"ohler and collaborators\cite{Hoehler}.

We can follow the same strategy to obtain a dispersive representation
of $D^+$ for nonzero values of $t$. This representation involves three
more subtraction constants, corresponding to the coefficients of $t$,
$\nu^2\, t$ and $t^2$ of the subthreshold expansion, as well as a
dispersion integral over the $t$-channel cut. After the
simplifications made in section \ref{sec:simp}, the $s$-channel
imaginary part is linear in $t$ and one needs an additional dispersion
integral, which supplies the contributions of the $s$- and $u$-channel
cut which are proportional to $t$. In the case of the amplitudes
$B^\pm$, the $s$-channel imaginary part is independent of $t$, so that
one function suffices to describe the $s$-channel cut contribution for
each of these. Furthermore the $t$-channel imaginary part of $B^+$
vanishes to the order of our calculation, so that we need a total of
nine functions of a single variable to account for all the low energy
singularities of the amplitude to $O(q^4)$.

\section{Conclusion}

We have presented a formulation for BCHPT (infrared regularization) in
which both, Lorentz invariance and the chiral power counting rules are
preserved at every stage of the calculation. The method relies on the
fact that for noninteger values of the regularization parameter $d$,
the infrared singular part of the relativistic loop integrals can be
unambiguously separated from the remainder. Since the remainder is
free of infrared singularities, it is a polynomial in the external
momenta and quark masses to any finite order of the chiral expansion
and can be absorbed into a redefinition of the coupling constants of
the effective Lagrangian.

We have shown that the chiral expansion of the $\pi N$-scattering
amplitude does not converge in the vicinity of the threshold at
$t=4\M^2$. The breakdown of the expansion is related to the fact that
the expansion of the nucleon propagator around the static limit
destroys some of the analytic properties of the amplitude. In the
relativistic formulation of the effective theory the correct
analytic structure is guaranteed and one obtains a representation,
which is valid in the entire low energy region.

The representation of a given amplitude obtained in the relativistic
framework contains in general also singularities outside the low
energy region. We have shown how to simplify the representation of the
$\pi N$-scattering amplitude at $O(q^4)$, by chirally expanding the
contributions of these singularities. As a result of these
simplifications, the scattering amplitude can be written as the Born
term, a polynomial containing the low energy constants and nine
functions of a single variable. The corresponding functions are given
in terms of dispersion integrals over the imaginary parts of the one
loop graphs.


\begin{thebibliography}{99}

\bibitem{hbchpt} E.~Jenkins and  A.~V.~Manohar, \Journal{{\em Phys.~Lett.~}B}{255}{558}{1991}.

 V.~Bernard, N.~Kaiser, J.~Kambor and U.-G. Meissner, \Journal{{\em Nucl.~Phys.}~B}{388}{315}{1992}.


\bibitem{gss}
  J.~Gasser, M.~E.~Sainio and A.~\v{S}varc, \Journal{{\em Nucl.~Phys.}~B}{307}{779}{1988}.
 
\bibitem{Becher:1999he}
T.~Becher and H.~Leutwyler,
%``Baryon chiral perturbation theory in manifestly Lorentz invariant form,''
\Journal{{\em Eur.\ Phys.\ J.}~C}{9}{643}{1999}

\bibitem{Tang} H.-B.~Tang, {\it A new approach to chiral perturbation theory
for matter fields}, hep-ph/9607436

P.~J.~Ellis and H.-B.~Tang, \Journal{{\em Phys.~Rev.}~C}{57}{3356}{1998}

\bibitem{julia} J.~Schweizer, {\em Low energy representation for
the axial form factor of the nucleon}, diploma thesis, Bern 2000.


\bibitem{Kubis}
B.~Kubis and U.~Meissner,
{\em Low energy analysis of the nucleon electromagnetic form factors}, 
hep-ph/0007056.
%%CITATION = HEP-PH 0007056;%%

B.~Kubis and U.~Meissner,
{\em Baryon form factors in chiral perturbation theory}, 
hep-ph/0010283.
%%CITATION = HEP-PH 0010283;%%

\bibitem{BL} T.~Becher and H.~Leutwyler, to appear.
\bibitem{Hoehler}
G.~H\"ohler, in {\em Landolt-B\"ornstein}, {\bf 9b2}, 
ed.~H.~Schopper (Springer, Berlin, 1983).


\end{thebibliography}
\end{document}